\documentclass[12pt,oneside,dvips]{article}
\usepackage{graphics}

\begin{document}
\pagenumbering{arabic}
\begin{center}
\begin{Large}
{\bf  Chaotic Monte Carlo computation: a dynamical
 effect of random-number generations }\\
\end{Large}
\end{center}
%\begin{large}
\begin{center}
   {\bf  Ken Umeno}\footnote{Electronic address: umeno@crl.go.jp} \\
   {\it Communications Research Laboratory}\\
{\it Ministry of Posts and Telecommunications}\\
   {\it 4-2-1 Nukui-Kitamachi, Koganei, Tokyo 184-8795, Japan}\\[1cm]
\end{center}
\begin{center}
{\bf Abstract}
\end{center}
{\bf 
Ergodic dynamical systems with absolutely continuous invariant probability measures 
are implemented as  random-number generators for Monte Carlo computation. 
Such chaos-based Monte Carlo computation yields sometimes unexpected 
dynamical dependency behavior which cannot be explained by the usual statistical argument. 
We resolve  the problem of its origin of this behavior by considering the effect of dynamical correlation on chaotic random-number generators.  
 Furthermore, 
we find that  superefficient 
 Monte Carlo computation can be carried out by using  
chaotic dynamical systems as random-number generators. Here 
superefficiency means that  the expectation 
value of the square of the error decreases to 0  as \(\frac{1}{N^{2}}\) 
 with \(N\) successive observations for \(N\rightarrow\infty\), 
whereas the conventional Monte Carlo simulation gives the square of the error as \(\frac{1}{N}\). The computation speed of the superefficient case 
does not depend on the dimensionality \(s\) of the problems and, hence, it is superior 
 to the low-discrepancy sequences yielding the square of the error 
\(\frac{(lnN)^{2s}}{N^{2}}\) in the known theoretical bounds.  By deriving  
a  necessary and sufficient condition for the superefficiency, 
it is shown  that   such  high-performance Monte Carlo simulations can be carried out only if  there exists a strong correlation of chaotic dynamical variables. Numerical 
calculation illustrates this dynamics dependency and the superefficiency of chaos Monte Carlo computations.}\\[1cm]
\begin{flushleft}
02.70.Lq 05.45.+b\\[1cm]
\end{flushleft}
\clearpage
\section{Introduction}
The applications of  physical processes to efficient computation  have 
been recently gaining attention. Among many kinds of natural phenomena, chaos
 has a peculiar merit such that while the implementations of chaos are rather easy in computers or in some physical devices, it exhibits  stochastic behavior as well as  the deterministic nature. Thus, it is of great interest to investigate what
kinds of efficient computations can be  harnessed by chaos.    
Since chaotic phenomena are described  by  deterministic dynamical systems, the numbers generated by chaotic dynamical systems can be considered as not purely random numbers but {\it pseudo-random} numbers.
So the right question must be focused 
on the distinguishable things between chaos and  the conventional 
pseudo random-number generators to perform some computation.
While various proposals of tests for {\it good} random-number generators have been  already  extensively 
studied  from the viewpoint of general aspects of  random-number generations\cite{knuth}, such comparison studies between chaos and pseudo random-number generators  have rarely been made. One exception is the study by  Phatak and Rao, which showed that the logistic map passed some statistical tests 
that ideal i.i.d.(independently, identically distributed) random-number generators 
must pass \cite{phatak}. In this paper, we study chaotic (ergodic) dynamical systems as   special-purpose random-number generators  for Monte Carlo methods, which are very powerful and general methods with many practical applications \cite{binder,fishman}. 

From the beginning, the key issue 
of the Monte Carlo  method has been the method of  generating random numbers \cite{metropolis}.  
However, no efficient chaotic random-number generator which is
 superior to  the conventional pseudo random-number generators
 in Monte Carlo simulations has yet reported.
The main purpose of this paper is to show
 that  {\it superefficient} Monte Carlo computation 
can be performed  by utilizing dynamical correlation of chaotic dynamical systems. 
In this superefficient case,  the expectation value of the square of the error
in Monte Carlo simulations decreases to  0 as \(\frac{1}{N^{2}}\) with
\(N\) successive observations for \(N\rightarrow \infty\), while the corresponding value of the conventional 
Monte Carlo simulations decreases to  0 as \(\frac{1}{N}\) for \(N\rightarrow \infty\).
The convergence speed of superefficient Monte Carlo simulations 
is therefore greatly accelerated by making use of chaotic dynamical systems as 
random-number generators.  
Since any  
Monte Carlo computation can be formulated as  a problem of evaluating  
an integral over a certain domain, here we consider the problem of 
Monte Carlo integration to elucidate the superefficiency of chaos-based  Monte Carlo computation. We do this by analyzing   
how a local dynamical characteristic (thus, non-random nature) of  chaotic 
random-number generators globally affects macroscopic (statistical) observable errors. 
  The Monte Carlo method is essentially based on the ergodic principle:  
When random numbers \(\{\xi_{j}\}_{j=1\cdots}\) are  uniformly 
distributed over \(\Omega\) (thus ergodic with respect to the Lebesgue measure on 
\(\Omega\)), the following identity   holds according to the ergodic 
theorem (or equivalently according to  the strong law of large numbers). 
\begin{equation}
\int_{\Omega}f(x)dx=\lim_{N\rightarrow \infty}\frac{1}{N}\sum_{ j=1}^{N}f(\xi_{j}).
\end{equation}  
A sum \(\frac{1}{N}\sum_{j=1}^{N}f(\xi_{j})\) can thus approximate the integral \(\int_{\Omega}f(x)dx\) with an error in the order of \(\frac{1}{\sqrt{N}}\) when  \(N\) is a large number.
In this paper, we combine this ergodic principle in Monte Carlo simulations  with  
some   ideas about generating  {\it nonuniform} random numbers by 
ergodic dynamical systems \cite{ulam,adler,ku7,ku8}.     
This paper is organized as follows. In Section 2, our Monte Carlo algorithm based on 
ergodic (chaotic) dynamical systems is explained. In Section 3, we demonstrate that 
phenomena of dynamical dependency of chaos-based Monte Carlo simulations are illustrated by a set of chaotic dynamical systems with  unique invariant measures. 
In Section 4, we clarify the necessary and sufficient condition for 
achieving the superefficiency such that the proposed  chaos-based Monte Carlo simulations are superior to  conventional Monte Carlo 
Simulations. Section 5 presents numerical results of such superefficient chaos-based Monte Carlo simulations    for a one-dimensional integration problem with integrands 
via Chebyshev-polynomial expansions. In Section 6, numerical results of superefficient chaos-based Monte Carlo simulations for a multi-dimensional integration problem are given. Section 7 presents  numerical results of superefficient chaos-based Monte Carlo simulations for an integration problem over the infinite support \((-\infty,\infty)\).   
A physical meaning of these superefficient chaos Monte Carlo computations 
is given in Section 8. And a summary and discussion are given in Section 9.

\section{Chaos-Based Monte Carlo Algorithm}
Consider a  dynamical system 
   \(X_{n+1}=F(X_{n})\) which is ergodic with respect to an  invariant 
probability measure \(\rho(x)dx\) with \(\rho(x)\) being a continuous density function 
\(\rho:\Omega\rightarrow R\).  This means that the measure \(d\mu(x)=\rho(x)dx\) 
is  invariant under time evolution and, furthermore, is absolutely continuous with 
respect to the Lebesgue measure on \(\Omega\). 
 This dynamical system can thus be seen as a random-number generator with a sampling measure \(\rho (x)dx  \)
on the domain \(\Omega\). 
In this case, according to the Birkhoff Ergodic Theorem \cite{birkhoff}, 
for any  function \(A(x)\) satisfying 
\begin{equation}
 \int_{\Omega}|A(x)|dx<\infty,
\end{equation}
the time average,   \(\lim_{N\rightarrow\infty}\frac{1}{N}\sum_{i=0}^{N-1}\frac{A(X_{i})}{\rho(X_{i})}\) is equal to the 
space average \(\int_{\Omega}A(x)dx\)  for almost every \(X_{0}\) with respect to 
the probability measure \(\rho(x)dx\) such that
\begin{equation}
\label{ergodicity}
  \overline{ A/\rho} = \langle A/\rho \rangle=\int_{\Omega}A(x)dx,
 \end{equation}
where 
\begin{equation}
\label{eq:ergordicity}
 \overline{B}
 \equiv
\lim_{N\rightarrow \infty}\frac{1}{N}\sum_{i=0}^{N-1}B(X_{i})
\end{equation}
 and 
\begin{equation}
 \langle B \rangle\equiv
 \int_{\Omega}B(x)\cdot \rho(x)dx.
\end{equation} 
 This means that 
the space average represented by \(\int_{\Omega}A(x)dx\) can be computed by the time-average of successively generated observables \(B_{i}\equiv B(X_{i})=A(X_{i})/\rho(X_{i})\)  by generated by this chaotic dynamical 
system. In the same way, we can obtain
a Monte Carlo algorithm for  multiple-dimensional integrals:
\begin{equation}
\overline{A/\rho_{s}}\equiv
\lim_{N\rightarrow\infty}\frac{1}{N}\sum_{i=0}^{N-1}\frac{A(X_{1,i},\cdots,X_{n,i})}{\prod_{j=1}^{s}\rho(X_{j,i})}=\langle A/\rho_{s}\rangle=
\int_{\Omega}A(x)dx,
\end{equation}
 where \(\rho_{s}(x)=\prod_{i=1}^{s}\rho(x_{i}),dx=\prod_{i=1}^{s}dx_{i}\), and 
\(\langle B(x) \rangle=\int_{\Omega}B(x)\rho_{s}(x)dx\). Here, we assume that 
each dynamics \(X_{j,i+1}=F_{j}(X_{j,i})\) has the 
same invariant probability measure 
 \(\rho(x)dx\).
Thus, any  ergodic dynamical systems  with respect to an explicit invariant probability 
measures \(\rho(x)dx \), in principle, can serve as random-number generators for Monte Carlo simulations in calculating arbitrary  multiple integrals. 
However,  such dynamical Monte Carlo computation cannot be carried out unless  
their invariant densities \(\rho(x)\) are explicitly known. In recent years, 
many chaotic dynamical systems have been  derived, which have explicit invariant  measures with continuous densities \cite{ku7,ku8}. Hence these chaotic dynamical systems
can offer their applications for  use  as random-number generators for Monte Carlo simulations.
Clearly, these random-number generators using chaos have  strict time-correlation which  is not desirable for ideal random-number generators. Suppose \(N\) successive observations, \(A_{i}\equiv A(X_{i}),\rho_{i}\equiv \rho(X_{i}),\quad i=1,\cdots,N\), of  quantities \(A(x)\) and a density function
\(\rho(x)\) have been stored.
We consider the expectation value of the square of the error
   as 
\begin{equation}
\label{eq:variance}
\sigma(N)\equiv  \langle\langle [\frac{1}{N} 
 \sum_{i=0}^{N-1} A_{i}/\rho_{i} -
 \langle A/\rho\rangle]^{2}   \rangle\rangle,
\end{equation}
where the expectation of \(B\) denoted by  \(\langle\langle B \rangle\rangle \) means
 an ensemble average with respect to the initial conditions \(X_{0}\) with  
a sampling measure \(\rho(x)dx\). As proved in Appendix A, this  expectation 
value \(\sigma(N)\) is  given   by the two-point 
correlation functions of \(B_{i}=B(X_{i})\) as follows: 
\begin{equation}
\label{eq:correlations}
\sigma(N) = \frac{1}{N}
\{\langle B^{2}\rangle -\langle B \rangle^{2}\}
 +\frac{2}{N}\sum_{j=1}^{N}(1-\frac{j}{N})\{\langle B_{0}B_{j}\rangle 
   -\langle B \rangle^{2}\},
\end{equation} 
where \( B(x)\equiv A(x)/\rho(x)\).
From Eq. (\ref{eq:correlations}), we can see that \(\sigma(N)\)
  is composed of   {\it the statistical 
variance term} 
\begin{equation}
\sigma_{s}(N)\equiv \frac{1}{N}
\{\langle B^{2}\rangle -\langle B \rangle^{2}\},
\end{equation}
 which purely depends on the form 
of the integrand \(B\)
 and {\it the dynamical correlation term} 
\begin{equation}
\sigma_{d}(N) \equiv \frac{2}{N}\sum_{j=1}^{N}(1-\frac{j}{N})\{\langle B_{0}B_{j}\rangle 
   -\langle B \rangle^{2}\},
\end{equation} 
which {\it depends} on the
 chaotic dynamical systems \(X_{n+1}=F(X_{n})\) utilized as random-number generators. 
From the above formula, 
when there exists a {\it negative} covariance(correlation) term such that
\begin{equation}
 \sum_{j=1}^{N}\{\langle B_{0}B_{j}\rangle-\langle B\rangle^{2}\}<0,
\end{equation}
 the total variance can be reduced. So such negative 
correlated variables ({\it antithetic variables}) have been used 
as a variance reduction technique in the Monte Carlo method\cite{hammersley}. 
In the next section, we consider such a 
 dynamical 
correlation term of chaotic random-number generators in  detail.
\section{Dynamical Dependency}
To test the dynamical correlation effect in chaotic random-number generators, 
we investigate chaos-based Monte Carlo computations of simple integration problems on the unit interval \(\Omega=[0,1]\)
by using chaotic dynamical systems with the same statistics described 
by a unique invariant measure. Chaotic dynamical systems utilized as random-number generators here are listed in Table 1.
In all of the numerical computations in this paper, we use \(M(=1000)\) different initial conditions \(X_{0}(j)\) for \(j=1,\cdots,M\). We measure the  numerical values of an empirical average of the square of the error
defined by 
\begin{equation}
V(N)=\frac{1}{M}\sum_{j=1}^{M}[\frac{1}{N}\sum_{i=0}^{N-1}A_{i}(j)/\rho_{i}(j)
 -\langle A/\rho\rangle]^{2}.
\end{equation}
Since the numerical sampling does not obey the probability measure \(\rho(x)dx\) exactly, the observed 
value \(V(N)\) is not equal to the exact ensemble average  of the square of the error \(\sigma(N)\) but fluctuates around 
\(\sigma(N)\). However, we can expect that 
\begin{equation}
   V(N)\rightarrow \sigma (N) \quad M\rightarrow \infty
\end{equation}
because the central limit theorem holds for random variables 
\begin{equation}
v_{j}\equiv [\frac{1}{N}\sum_{i=0}^{N-1}A_{i}(j)/\rho_{i}(j)-\langle A/\rho\rangle]^{2},\quad j=1,\cdots,M
\end{equation}
 for \(M\rightarrow \infty\).
Thus, the simulation results here  does not depend on the precise sampling information of the initial 
data \(\{X_{0}(j)\}_{j=1,\cdots,M}\).
Figure 1 shows that although Ulam-von Neumann map,  Cubic map and  Quintic map 
have the same invariant measure \(\rho(x)dx=\frac{dx}{\pi\sqrt{x(1-x)}}\), the resulting error variances of chaos-based Monte Carlo integration  of \(I=\int_{0}^{1}x^{3}dx=\frac{1}{4}\) exhibit discrepancy between them. Similar results hold for the generalized Ulam-von Neumann map and generalized Cubic map, both of which
have 
the same invariant measure \(\rho(x)dx=\frac{dx}{\pi\sqrt{x(1-x)(1-\frac{1}{2}x)(1-\frac{3}{10}x)}}\).
 Figure 2  shows that there exists a significant dynamical dependency of the 
expectation value of the square of the errors for chaos-based Monte Carlo integration 
of \(I=\int_{0}^{1}x(1-x)dx=\frac{1}{6}\) by using the same family  of chaotic 
dynamical systems  of Fig.1. Furthermore, these figures show that 
the chaotic random-number generator resulting in the most efficient computation 
depends on the integrand(the generalized Cubic map for Fig. 1 and Ulam-von Neumann map for Fig. 2). 
Therefore, it is clear that these numerical results of chaotic random-number generators cannot  
 be explained by the usual statistical argument such as the importance sampling. 
Hence, we must consider
the dynamical effect of these chaotic random-number generators to explain
these dynamical dependency phenomena. In the next section, we consider an
extreme case of dynamical dependency resulting in superefficient 
 chaos-based Monte Carlo simulations. 
\section{Condition for Superefficiency}
Here, we derive  the condition for attaining the superefficiency 
of chaos-based Monte Carlo simulations such that the expectation value 
of the square 
of the error decreases to 0 as \(\frac{1}{N^{2}}\) for \(N\rightarrow\infty\). 
As mentioned  in Section 2, if random variables  generated by 
chaotic random-number generators have a negative correlation, the resulting expectation value  \(\sigma(N)\) of the square of the error 
 can be reduced as the usual variance reduction technique.  
Furthermore, we have a stronger statement in the case that a chaotic dynamical 
system has a mixing (thus, ergodic) property as follows:\\
\newtheorem{theorem}{Theorem}
\begin{theorem}[Superefficiency Condition]
\label{theorem:1}
 The expectation value of the square of the error \(\sigma(N)\) 
decreases to zero as \(\frac{1}{N^{2}}\) as 
\begin{equation}
\label{eq:super_corr}
\sigma(N)=-\frac{2}{N^{2}}\sum_{j=1}^{N}j\{\langle B_{0}B_{j}\rangle-\langle B\rangle^{2}\}=O(\frac{1}{N^{2}})\rightarrow 0\quad \mbox{for}\quad N\rightarrow \infty,
\end{equation}
if and only if the relation
\begin{equation}
\label{eq:super_condition}
 \langle B^{2}\rangle -\langle B\rangle^{2}+2
\sum_{j=1}^{\infty}\{\langle B_{0}B_{j}\rangle-\langle B\rangle^{2}\}=0 
\end{equation} 
is satisfied.
\end{theorem} 
The proof of the theorem is given as follows.  Since  \(\sigma (N)\) has the following 
identity from formula (\ref{eq:correlations}): 
\begin{equation}
 \sigma(N)=\frac{  \langle B^{2}\rangle -\langle B\rangle^{2}+2\sum_{j=1}^{N}\{\langle B_{0}B_{j}\rangle-\langle B\rangle^{2}\} }{N}-\frac{2}{N^{2}}\sum_{j=1}^{N}j\{\langle B_{0}B_{j}\rangle-\langle B\rangle^{2}\},
\end{equation}
we have the relation 
\begin{equation}
  \sigma(N)=O(\frac{1}{N^{2}})\quad \mbox{for}\quad N\rightarrow \infty\iff  \langle B^{2}\rangle -\langle B\rangle^{2}+2\sum_{j=1}^{\infty}\{\langle B_{0}B_{j}\rangle-\langle B\rangle^{2}\}=0.
\end{equation}
According to the mixing property of the chaotic dynamical systems with Lyapunov exponents \(\lambda(>0)\), we have 
\begin{equation}
\langle B_{0}B_{j}\rangle-\langle B\rangle^{2}=O[e^{-j\lambda}]\rightarrow 0,
\quad \mbox{for}\quad j\rightarrow \infty,
\end{equation}
which assures the convergence of the infinite sums \(\sum_{j=1}^{\infty}\{\langle B_{0}B_{j}\rangle-\langle B\rangle^{2}\}\) and 
\(\sum_{j=1}^{\infty}j\{\langle B_{0}B_{j}\rangle-\langle B\rangle^{2}\}\) in the 
superefficiency condition (\ref{eq:super_condition}). 
It is clear that 
Theorem 1 gives us a necessary and sufficient condition for the superefficiency  of the chaos-based Monte Carlo simulations. Furthermore, Theorem 1  about the 
condition for  superefficiency does not depend on the dimensionality \(s\) of integration domains.
Thus, we call 
the condition given in Eq.(\ref{eq:super_condition}) the {\it superefficiency condition} for general \(s\)-dimensional Monte Carlo integration problems. 

The   superefficiency here  
is reminiscent of {\it low-discrepancy sequences} (or {\it quasi-random numbers}) generated by a deterministic algorithm,  where the theoretical 
error bounds are known to be of size \(O[\frac{(\mbox{ln}N)^{s}}{N}]\) for \(s\)-dimensional integration 
problems, which are also superior to the conventional Monte Carlo  simulations with the error of size \(O(\frac{1}{\sqrt{N}})\)\cite{niederreiter}. 
Consequently, the theoretical values of the square of the error  in the low-discrepancy sequences 
are in the order 
of \(O[\frac{(\mbox{ln}N)^{2s}}{N^{2}}]\). Thus, 
a particular superefficient Monte Caro simulation using 
chaos is superior to  low-discrepancy sequences in the sense that 
the expected 
value \(\sigma(N)\) of the square of the error in the  superefficient Monte Carlo simulation 
is \(O(\frac{1}{N^{2}})\) where the logarithmic factor \((\mbox{ln}N)^{2s}\), which is not  negligible for a large dimensionality \(s\), is removed from the corresponding order of the low-discrepancy sequences. Table 2 compares  the data about the order of the square of the errors. 
Besides the differences in the order, 
the following differences  between the two deterministic methods of Monte Carlo computations exist. 
While  it is difficult for explicitly estimating   
the error  for low-discrepancy sequences, this superefficient Monte Carlo 
computations can give an exact estimate of the mean error variance as in Eq. (\ref{eq:super_corr}). However,   
the present superefficient chaos-based Monte Carlo computations, which are superior to 
low-discrepancy sequences
in higher dimension \(s\), can be carried out only if  
 there exists a negative 
correlation of chaotic dynamical variables satisfying superefficiency condition 
(\ref{eq:super_condition}).
In the following sections, we will present concrete examples yielding superefficient 
chaos Monte Carlo computations.

\section{Superefficiency: A Case of Chebyshev Maps}
Here, we  achieve superefficiency \(\sigma(N)=O(\frac{1}{N^{2}})\)
 for a certain family of one-dimensional integration problems by
 using   a set of specific chaotic   dynamical systems \(X_{n+1}=T_{p}(X_{n})\), where  \(T_{p}(X)\)
is the \(p\)-th order Chebyshev polynomial defined  by 
 \(T_{p}[\cos (\theta)]\equiv \cos (p\theta)\) at \(p\geq 2\).  Examples of 
Chebyshev polynomials are given by 
\begin{equation}
 T_{0}(X)=1,T_{1}(X)=X,T_{2}(X)=2X^{2}-1,T_{3}(X)=4X^{3}-3X,\cdots.
\end{equation}
 It is known \cite{adler} that  
these  Chebyshev maps \(T_{p}\) are mixing (thus, ergodic)  with respect to the invariant measure  \(\frac{dx}{\pi \sqrt{1-x^{2}}}\)
on the domain \(\Omega=[-1,1]\) for \(p\geq  2\) and they have 
Lyapunov exponents \(\mbox{ln} p\). 
Note that the chaotic dynamical system  \(X_{n+1}=T_{2}(X_{n})=2X_{n}^{2}-1\) 
is equivalent to the well-known logistic dynamical system \(Y_{n+1}=4Y_{n}(1-Y_{n})\)  with an invariant measure \(\frac{dy}{\pi
\sqrt{y(1-y)}}\) on the unit interval \([0,1]\)\cite{ulam}.  In the same way,  chaotic  dynamics caused by the third-order
Chebyshev map \(X_{n+1}=T_{3}(X_{n})=4X_{n}^{3}-3X_{n}\) is equivalent to the cubic 
chaos  \(Y_{n+1}=Y_{n}(3-4Y_{n})^{2}\)   with Lyapunov exponent \(\mbox{ln}3\). It is also known that a system of Chebyshev polynomials 
constitutes a complete orthonormal system  satisfying the relations 
\begin{equation}
\label{eq:che_orthonormal}
\int_{-1}^{1}T_{i}(x)T_{j}(x)\rho(x)dx=\delta_{i,j}\frac{(1+\delta_{i,0})}{2},
\end{equation}
 where 
\(\delta_{i,j}\) stands for the  Kronecker delta function such that
\renewcommand\displayindent{3zw} 
\begin{eqnarray}
\delta_{i,j}=
\left\{\begin{array}{ll}
1 & i=j \\
0 & i\ne j.
\end{array}
\right.
\end{eqnarray}  
 Consider a one-dimensional continuous function  \(B(x)\) as an integrand
 on the domain \(\Omega=[-1,1]\). By Weierstrass's approximation theorem,  for an arbitrary \(\epsilon>0\), there exists a polynomial function 
\(P(x)\) satisfying 
\begin{eqnarray}
\nonumber
 |B(x)-P(x)|<\epsilon \quad \mbox{for}\quad -1\leq x\leq 1.
\end{eqnarray} 
We can thus uniquely expand an arbitrary integrand (continuous function) \(B(x)\)
 in terms of Chebyshev polynomials (orthogonal-basis functions):
\begin{eqnarray}
\nonumber  
   B(x)=\sum_{k=0}^{\infty}b_{k}T_{k}(x)=\sum_{k=0}^{\infty}\frac{2}{1+\delta_{k,0}}\langle BT_{k}\rangle T_{k}(x), 
\end{eqnarray}
 where the coefficients are given by the formula
\begin{eqnarray}
\nonumber
b_{k}=\frac{2}{(1+\delta_{k,0})}\int_{-1}^{1}B(x)T_{k}\rho(x)dx
 =\frac{2}{1+\delta_{k,0}}\langle BT_{k} \rangle.
\end{eqnarray}
The completeness of the system of  Chebyshev orthogonal functions
assures that 
\begin{equation}
   b_{k}\rightarrow 0\quad \mbox{for}\quad k\rightarrow \infty.
\end{equation}
Thus, it can be said that any continuous function 
can be well approximated in terms of  finite numbers  of Chebyshev polynomials.
So we consider the Monte Carlo integration problem of an integrand 
\begin{equation}
\label{eq:chebyshev_expand}
B(x)=\sum_{k=0}^{L}b_{k}T_{k}(x).
\end{equation}
 Since
\begin{equation}
B[T_{j}(x)]=\sum_{k=0}^{L}b_{k}T_{k}[T_{j}(x)]=\sum_{k=0}^{L}
 b_{k}T_{kj}(x),
\end{equation}
we can calculate two-point correlation functions \(\langle B_{0}B_{j} \rangle
(=\langle B(x)B[T_{p^{j}}(x)] \rangle) \) computed by the \(p\)-th Chebyshev map \(T_{p}\)
as follows:
\renewcommand\displayindent{3zw}
\begin{eqnarray}
\label{eq:cheby_corr}
   \langle B_{0}B_{j} \rangle &=& \langle b_{0}+\sum_{k=1}^{L} b_{k}T_{k}(x),
   b_{0}+\sum_{k=1}^{L}b_{k}T_{k\cdot p^{j}}(x) \rangle \nonumber\\
   &=& \left\{ \begin{array}{ll} 
b_{0}^{2}+\frac{1}{2}\sum_{k=1}^{[L/p^{j}]}b_{kp^{j}}b_{k} & j\leq[\log_{p}L] \\
b_{0}^{2} & 
j\geq [\log_{p}L]+1 
\end{array}
\right. 
\end{eqnarray}
where \([x]\) is the largest non-negative  integer part of \(x\).
It can be seen from Eq. (\ref{eq:cheby_corr}) that when 
\(p>L\), 
 the \(p\)-th Chebyshev map causes no dynamical effect such that
   \(\sigma(N)=\sigma_{s}(N)\).
Hence from now on, we consider the case  \(p\leq L\). 
The statistical variance term is given by 
\begin{equation}
\sigma_{s}(N)= \frac{1}{N}\{\langle B^{2} \rangle-\langle B\rangle^{2}\}=
\frac{1}{2N}(\sum_{m=1}^{L}b^{2}_{m}),           
\end{equation}
 and
the dynamical correlation term of the variance of error is given by 
\begin{equation}
\label{eq:cheby_dyna}
   \sigma_{d}(N)=\frac{2}{N}\sum_{j=1}^{N}(1-\frac{j}{N})\{\langle B_{0}B_{j}\rangle -\langle B\rangle^{2}\}=
\frac{2}{N}\sum_{j=1}^{[\log_{p}L]}
  \{(1-\frac{j}{N})\}\sum_{m=1}^{[L/p^{j}]}b_{mp^{j}}b_{m}.
\end{equation} 

The  condition for the superefficiency in Eq. (\ref{eq:super_condition})
 is thus given  in terms of the coefficients of the Chebyshev expansions of an integrand
\(B(x)=\sum_{m=0}^{L}b_{m}T_{m}(x)\) as follows:
\begin{equation}
\label{eq:super_condition2}
\sum_{j=1}^{\infty}\{\langle B_{0}B_{j}\rangle -\langle B\rangle^{2}\}
=
\sum_{j=1}^{[\log_{p}L]}\sum_{m=1}^{[L/p^{j}]}b_{mp^{j}}b_{m}
=-\frac{1}{2}(\sum_{m=1}^{L}b_{m}^{2})<0,
\end{equation}
where the Monte Carlo simulations are computed by 
the \(p\)-th Chebyshev map \(T_{p}\). 
Therefore, for any integrand given by Chebyshev expansions as Eq. 
(\ref{eq:chebyshev_expand}),  
a dynamical correlation term represented by  the L.H.S. of 
Eq. (\ref{eq:super_condition2}) must thus be {\it negative} to satisfy the superefficiency 
condition
 As is clearly seen in Eq. (\ref{eq:super_condition2}), the constant 
term \(b_{0}\) is {\it irrelevant} for the superefficiency. 
 Here we can give a family of examples that meet the superefficiency condition as follows. 
 
Consider the case that a normalized integrand  \(B(x)=\frac{A(x)}{\rho(x)}\) has  the following  form
\begin{eqnarray}
\label{eq:one_integrand}
B(x)=a_{c}+\sum_{m=0}^{L}a_{m}T_{p^{m}}(x)=\frac{A(x)}{\rho(x)},
\end{eqnarray}
where \(p\geq 2\).
Note that  
\begin{equation}
\langle B \rangle=a_{c}\quad \mbox{and}\quad  
\langle B^{2} \rangle=a_{c}^{2}+\frac{1}{2}\sum_{m=0}^{L}a^{2}_{m}.
\end{equation}

In this case, the following two-point correlation functions are simply given by  
\begin{equation}
\langle B_{0}B_{l}\rangle=\langle a_{c}+\sum_{m=0}^{L}a_{m}T_{p^{m}}(x),
a_{c}+\sum_{m=0}^{L}a_{m}T_{p^{m+l}}(x) \rangle=
a_{c}^{2}+\frac{1}{2}\sum_{m=l}^{L}a_{m}a_{m-l},
\end{equation}
 Thus, with the use of formula (\ref{eq:correlations}), the expectation value of the square of  the error \(\sigma(N)\) is explicitly   given  by the following formula:
\begin{equation}
\label{eq:super1}
\sigma(N)=\frac{1}{2N}(\sum_{m=0}^{L}a_{m})^{2}
-\frac{1}{N^{2}}\sum_{l=1}^{L}\sum_{m=l}^{L}l a_{m}a_{m-l}.
\end{equation}
Therefore, superefficiency condition (\ref{eq:super_condition}) is equivalent to 
the simple condition
\begin{equation}
\label{eq:super_2}
  \sum_{m=0}^{L}a_{m}=0.
\end{equation} 
In other words, when condition  (\ref{eq:super_2}) is satisfied,
the \(O(1/N)\) term of \(\sigma (N)\)
 can be eliminated.  Hence, if an integrand has a form of 
Eq. (\ref{eq:one_integrand}) with the condition \(\sum_{m=0}^{L}a_{m}=0\), 
the superefficient Monte Carlo computation can be carried out by the \(p\)-th 
order Chebyshev chaos map such that the resulting 
expectation value of the square of the error are given 
as follows:
\begin{equation}
\label{eq:super_sigma_c}
\sigma(N)=-\frac{1}{N^{2}}\sum_{l=1}^{L}\sum_{m=l}^{L}la_{m}a_{m-l}=O(\frac{1}{N^{2}}).
\end{equation}
On the other hand, even if an integrand satisfies the above condition,
 other types of chaotic dynamics, such as,  
\(X_{j+1}=T_{p'}(X_{j})\) for \(p'\ne p\) then cause 
 the time-correlation functions to exhibit  {\it no} dynamical  effect:
\begin{equation}
\langle B_{0}B_{l}\rangle =\langle a_{c}+\sum_{m=0}^{L}a_{m}T_{p^{m}}(x),
a_{c}+\sum_{m=0}^{L}a_{m}T_{p^{m}\cdot p'^{l}}(x) \rangle =a_{c}^{2}=
\langle B\rangle^{2}.
\end{equation}
Therefore, \(\sigma(N)\) in this case does not so 
quickly  decrease to 0 for \(N\rightarrow\infty\) as the superefficient cases but normally converges to 0:
\begin{equation}
  \sigma(N)=\sigma_{s}(N)=\frac{1}{2N}(\sum_{m=0}^{L}a_{m}^{2})=O(\frac{1}{N}).
\end{equation}
This means that   specific chaotic (Chebyshev) dynamical systems do exist in all of the Chebyshev chaotic dynamical systems, which 
make Monte Carlo computation {\it superefficient}  and this selection of the Chebyshev chaotic dynamical systems for the superefficiency depends on 
a form of integrand, irrespective of the fact that all of the Chebyshev chaotic dynamical systems has the same invariant probability measure (the same statistics).
It   can thus be said that the superefficiency of chaos-based Monte Carlo simulations is caused not by a
statistical effect but  by a
 {\it  purely  dynamical effect} (thus, a  non-random effect) with a strong correlation of chaotic variables in random-number generators.
To numerically confirm 
 the interplay between an integrand
and chaotic dynamical systems chosen as random-number generators, 
 we consider a simple integrand on \(\Omega=[-1,1]\) as follows:
\begin{equation}
 A(x)=\frac{x-2x^{2}+1}{\pi\sqrt{1-x^{2}}}=[T_{1}(x)-T_{2}(x)]\rho(x)=B(x)\rho(x).
\end{equation}
 Here, this integrand corresponds to the case that \(a_{0}=1,a_{1}=-1,L=1\), and \(p=2\) in Eq. (\ref{eq:one_integrand}), which clearly satisfies the above condition of superefficiency.  

Thus, the random numbers generated by the second-order Chebyshev maps \(X_{n+1}=T_{2}(X_{n})\) are predicted to yield the  expectation value of the square 
of the error,
\begin{equation}
V_{T_{2}}(N)\approx \sigma_{T_{2}}(N)=\frac{1}{N^{2}}
\end{equation}
by the formula (\ref{eq:super1}), while the other types of the 
\(p\)-th  Chebyshev maps  at \(p\geq 3\) are predicted to  give the expectation value of square of error in the order of 
\begin{equation}
V_{T_{p}}(N)\approx \sigma_{T_{p}}(N)=\frac{1}{N}\quad \mbox{for}\quad p>2.
\end{equation}
Figure 2 shows that numerical results coincide with our prediction  
about the occurrence of the superefficiency  
 of  chaos-based Monte Carlo simulations.
The superefficiency achieved  here remarkably 
  contrasts 
with the conventional Monte Carlo simulations with
\(V(N)\approx \sigma(N)= O(\frac{1}{N})\) 
computed by  the other \(p\)-th Chebyshev maps \(p>2\). 
If \(p\) in Eq. (\ref{eq:one_integrand}) is a composite number  such as \(p=p'^{k}\) for an integer 
\(k(\geq 2)\), the \(p'\)-th order 
Chebyshev map \(T_{p'}\) also gives a superefficiency. To check this numerically,
 we consider an integrand 
\begin{eqnarray}
\nonumber
A(x)=[T_{1}(x)-T_{4}(x)]\rho(x)=\frac{(-8x^{4}+8x^{2}+x-1)}{\pi\sqrt{1-x^{2}}}. 
\end{eqnarray}
 In this case,   both the second-order Chebyshev map \(T_{2}\)
and the  fourth-order Chebyshev map \(T_{4}\) are predicted to 
give  
superefficient results such as
\begin{equation} 
V_{T_{2}}(N)\approx \sigma_{T_{2}}(N)=\frac{2}{N^{2}}
\end{equation}
 and 
\begin{equation}
 V_{T_{4}}(N)\approx \sigma_{T_{4}}(N)=\frac{1}{N^{2}}
\end{equation}
 respectively,
 while the third-order Chebyshev map \(T_{3}\) and the fifth-order 
Chebyshev map \(T_{5}\) are predicted to give the normal behavior of 
the expectation value of the square of the error 
\begin{equation}
V_{T_{p}}(N)\approx \sigma_{T_{p}}(N)=\frac{1}{N}\quad \mbox{for}\quad p=3,5.
\end{equation}
Figure 3 confirms that numerical data coincide with our prediction about the selection of 
chaotic dynamical systems  for  performing superefficient chaos-based Monte Carlo computations.  The sampling dependency of superefficiency were also numerically 
tested. Figure 4(a) shows the numerically obtained 
 error variance \(V(N)\) coincides well with  the exact mean value of the square of the error \(\sigma(N)\) in Eq. (\ref{eq:super_sigma_c})
for the superefficient Monte Carlo computation of an \(B(x)=T_{1}-T_{4}+T_{0}\) with the sampling measure well approximated by the invariant measure \(\rho(x)\).
Figure 4(b) shows that while there exists  discrepancy between the numerical value and the exact value of the mean square of the error for the same problem 
 with uniform sampling measure of initial data, the same kinds of superefficient 
Monte Carlo simulations are carried out. Thus, superefficient chaos-based Monte Carlo 
computation is very robust under our choice of the initial conditions of 
chaotic dynamical systems.  
\section{Superefficiency: A Case of Multi-Dimensional Integration}
As  for the case of one-dimensional Monte Carlo integration problems, we 
try to achieve the same superefficiency in the 
  multi-dimensional integration problem. 
Let us consider the domain of integration as  the \(s\)-dimensional cubic
\(\Omega_{s}=[-1,1]^{s}\). Even in this multi-dimensional case, there exists the 
multi-dimensional version of  Weierstrass's approximation theorem \cite{courant} 
which guarantees that  for an arbitrary \(\epsilon\), there exists a polynomial 
function \(P(x)=P(x_{1},\cdots,x_{s})\) such that 
\begin{equation}
 |B(x)-P(x)|<\epsilon \quad \mbox{for}\quad x\in \Omega_{s},
\end{equation}
where \(B(x)=B(x_{1},\cdots,x_{s})\) is a continuous function over the domain \(\Omega_{s}\). 
From the complete orthonormal property of the Chebyshev polynomials \(\{T_{l}(x)\}\),
we can construct \(T_{j_{1}}(x_{1})\cdots T_{j_{s}}(x_{s})\) as an element 
the  complete orthonormal basis functions spanning the  functional space \(\cal L\mit^{2}(\Omega_{s},\rho_{s})\) such that 
\begin{equation}
\cal L\mit^{2}(\Omega_{s},\rho_{s})=
\{B(x_{1},\cdots,x_{s})|\int_{\Omega_{s}}|B(x_{1},\cdots,x_{s})|^{2}\rho_{s}(x)
dx_{1}\cdots dx_{s}<\infty\}.
\end{equation}
Any continuous function \(B(x_{1},\cdots,x_{n}) \in \cal L\mit^{2}(\Omega_{s},\rho_{s}) \) can thus be uniquely expanded in terms of the products of the Chebyshev polynomials as follows:
\begin{equation}
   B(x_{1},\cdots,x_{s})=\sum_{j_{1},\cdots,j_{s}\geq 0}b_{j_{1},\cdots,j_{s}}T_{j_{1}}(x_{1})\cdots T_{j_{s}}(x_{s}).
\end{equation}
Therefore, as is the one-dimensional case, superefficiency condition are given 
for any continuous function \(B(x_{1},\cdots,x_{n})\in \cal L\mit^{2}(\Omega_{s},\rho_{s})\) in terms of coefficients of 
the Chebyshev expansions.
Here, to present multi-dimensional superefficiency,  we consider a more specific family of integrands with the form,
\begin{equation}
\label{eq:multi_cheby}
\frac{A(x)}{\rho_{s}(x)}=B(x)=B(x_{1},\cdots,x_{s})=a_{c}+\sum_{m=0}^{L}a_{m}T_{p_{1}^{m}}(x_{1})
\cdots T_{p_{s}^{m}}(x_{s}),
\end{equation}
where each dynamical variable \(X_{i,j}\) associated with \(x_{i}\) is computed 
by the \(p_{i}(\geq 2)\)-th order Chebyshev map as \(X_{i,j+1}=T_{p_{i}}(X_{i,j})\). 
Note that a  polynomial   
\(T(x_{1},\cdots,x_{n})= T_{p_{1}^{m_{1}}}(x_{1})\cdots T_{p_{s}^{m_{s}}}(x_{s})\) for \(p_{i}\geq 0,m_{i}\geq 0\) is also an element of orthonormal basis-functions which constitutes a 
 complete orthonormal system in the \(s\)-dimensional functional space \(\cal L\mit^{2}(\Omega_{s},\rho_{s})\).
  As in the one-dimensional case, by using the orthonormal property of the Chebyshev polynomials, we can  compute the following two-point correlation 
functions:
\begin{equation}
 \langle B_{0} B_{l} \rangle =a_{c}^{2}+\frac{1}{2^{s}}\sum_{m=l}^{L}a_{m}a_{m-l}.
\end{equation}
As will be proven in Appendix B,
the expectation value of the square of the error \(\sigma(N)\) is thus given by 
\begin{equation}
\label{eq:s_dimensional_sigma}
 \sigma(N)=\frac{1}{2^{s}N}(\sum_{m=0}^{L}a_{m})^{2}-\frac{1}{2^{s-1}N^{2}}
\sum_{l=1}^{L}\sum_{m=l}^{L}la_{m}a_{m-l}.
\end{equation}
Therefore, if an integrand satisfies the superefficiency condition \(\sum_{m=0}^{L}a_{m}=0\), which is equivalent to the relation (\ref{eq:super_condition}), 
\(s\) chaotic dynamical systems \(X_{i,j+1}=T_{p_{i}}(X_{i,j})\) for \(i=1,\cdots,s\) again yield  a superefficient Monte Carlo computation 
with the expectation value of the square of error 
in the  order of \(\frac{1}{N^{2}}\) as follows:
\begin{equation}
 \sigma(N)=-\frac{1}{2^{s-1}N^{2}}\sum_{l=1}^{L}\sum_{m=l}^{L}la_{m}a_{m-l}=O(\frac{1}{N^{2}}).
\end{equation}

 Thus, the superefficiency of chaos-based 
Monte Carlo computation can be carried out even for   multi-dimensional 
integration problems. 

To numerically check the multi-dimensional superefficiency of 
chaos-based Monte Carlo simulations, we consider a 
two-dimensional integrand as follows:
\begin{eqnarray}
\nonumber 
B(x,y)=xy-(2x^{2}-1)(4y^{3}-3y)=T_{1}(x)T_{1}(y)-T_{2}(x)T_{3}(y).
\end{eqnarray}
This integrand corresponds to \(a_{c}=0,a_{0}=1,a_{1}=-1,p_{1}=2,p_{2}=3\) and \(L=1\) 
in Eq. (\ref{eq:multi_cheby}). 
Thus, the superefficiency condition is satisfied. 
Therefore, if we use the chaotic dynamical systems \(X_{j+1}=T_{2}(X_{j})\) and \(Y_{j+1}=T_{3}(Y_{j})\) as random-number generators for \(x\) and \(y\) respectively, superefficient Monte Carlo computation is predicted to  be carried 
 out such that the numerically obtained expectation value of the square of the error  
\(V(N)\) is given by 
\begin{equation}
 V_{T_{2},T_{3}}(N) \approx \sigma_{T_{2},T_{3}}(N)= \frac{1}{2N^{2}},
\end{equation}
while the other chosen sets of 
chaotic dynamical systems yield  
\begin{equation}
V_{T_{p1},T_{p2}}(N)\approx \sigma_{T_{p1},T_{p2}}(N)=\frac{1}{2N}
\end{equation}
for \((p1,p2)=(2,2),(3,2),\mbox{or},(3,3)\).
Figure 5  shows that the numerical data coincide well with our theory about
the multi-dimensional superefficiency of chaos-based Monte Carlo simulations.
\section{Superefficiency: A Case of Non-Gaussian Chaos}
Let us consider an integration problem over the infinite support 
\(\Omega_{I}=(-\infty,\infty)\) with density function \(\rho_{NG}(x)\). In this case,  Weierstrass's approximation theorem cannot be directly applied.
However,
when we consider  
a functional space    
\begin{equation}
  \cal L\mit^{2} (\Omega_{I},\rho_{NG})=  \{f(x)| \int_{\Omega_{I}}|f(x)|^{2}\rho_{NG}(x)dx<\infty\},
\end{equation}
the chaos-based Monte Carlo integration of 
an  integrand \(f(x)\in \cal L\mit^{2} (\Omega_{I},\rho_{NG})\) 
can also be performed without any modification of the above algorithm. 
The Monte Carlo integration over the infinite support also has many applications 
such as the pricing of exotic options in financial markets. Furthermore, the problem of Monte Carlo integration over the infinite support is very important from the theoretical point of view of Monte Carlo computation, because 
we cannot directly use the usual uniform random-number generators but must consider 
some non-uniform random numbers over \(\Omega_{I}\). Usually, 
a procedure of Monte Carlo integration over the infinite support must include 
a transformation from the uniformly distributed random numbers to random numbers 
with non-uniform distributions such as the inverse method\cite{devroye}.
 Here without resort to a transformation of random variables, 
we  directly use certain ergodic  dynamical 
systems with invariant probability measures over  
the infinite support \(\Omega_{I}=(-\infty,\infty)\) as  random-numbers generators over \(\Omega_{I}\). 
Such ergodic dynamical systems over the infinite support 
were systematically  found from the multiplication formulas of \(\tan(\theta)\) and 
its related functions \cite{ku9}.  We can thus use these chaotic dynamical systems 
for this chaos-based Monte Carlo simulations over the 
infinite support \(\Omega_{I}\). Let us briefly explain ergodic mappings over \(\Omega_{I}\).
As will be shown in Appendices C and D,
by using  the topological conjugacy relation with  Chebyshev mappings \(T_{l}\), 
 we can derive an infinite number of ergodic transformations
\begin{equation}
 F_{l}(y)=h^{-1}\circ T_{l} \circ h(y), 
\end{equation}
 where \(h\) is a differential onto-mapping (diffeomorphism) such that 
\(h:(-\infty,\infty)\rightarrow (-1,1)\) is given by \(h(y)=\frac{1}{\sqrt{1+|y|^{2\alpha}}}\mbox{sgn}(y)\)  for \(\alpha>0\). 
These ergodic mappings \(\{F_{l}\}\) have 
 the same invariant probability measure \cite{ku9} 
\begin{equation}
\label{eq:nong_density}
  \rho_{NG}(y)=\frac{\alpha |y|^{\alpha-1}}{\pi(1+|y|^{2\alpha})}.
\end{equation} 
Furthermore, 
as will be shown in Appendix C,
we can consider the corresponding  orthonormal system of functions \(\{ P_{l}(y)\equiv T_{l}\circ h(y)\}\) satisfying the same orthogonal 
relation as the Chebyshev orthogonal polynomials as follows:
\begin{equation}
\int_{-\infty}^{\infty} P_{i}(y) P_{j}(y) \rho_{NG}(y)dy
=\delta_{i,j}\frac{(1+\delta_{i,0})}{2}.
\end{equation}
These orthogonal functions constitute a complete orthonormal system of functions 
in \(\cal L\mit^{2}(\Omega_{I},\rho_{NG})\).  
Hence, as in constructing multidimensional integrals in terms of Chebyshev polynomials where superefficiency condition
is attained, chaos-based Monte Carlo simulations for a family of integrands 
\begin{equation}
  B(x_{1},\cdots,x_{s})=a_{c}+\sum_{m=0}^{L}a_{m} P_{p_{1}^{m}}(x_{1})\cdots P_{p_{s}^{m}}(x_{s}),
\end{equation}
where an each dynamical variable \(X_{i,j}\) associated with \(x_{i}\) is computed by \(X_{i,j+1}=F_{p_{i}}(X_{i,j})\), are superefficient if the condition 
\begin{equation}
   \sum_{m=0}^{L}a_{m}=0
\end{equation}
is satisfied. 
Table 3 lists these  transformations \(\{F_{l}\}_{l=1,2,3}\) and their dual 
transformations 
\(\{F^{*}_{l}\}_{l=1,2,3}\) over \(\Omega_{I}\) and their related 
orthogonal functions \(\{P_{l}(x)\}_{l=1,2,3}\) and \(\{P^{*}_{l}(x)\}_{l=1,2,3}\) which were used for numerical simulations to attain the superefficiency over the infinite support. 
Figure 6(a) shows the graphs of ergodic transformations at \(\alpha=1\). Figure 
6(b) shows the graphs of the density functions \(\rho_{NG}(x)\) in 
Eq. (\ref{eq:nong_density}) over \(\Omega_{I}\) at several different \(\alpha\).  
In Fig. 7, we show that superefficient Monte Carlo computations of one-dimensional integrands \(B(x)=P_{2}(x)-P_{1}(x)=\frac{1-x^{2}}{1+x^{2}}-\frac{\mbox{sgn}(x)}{\sqrt{1+x^{2}}}\) and 
\(B(x)=P^{*}_{2}(x)-P^{*}_{1}(x)=\frac{x^{2}-1}{x^{2}+1}-\frac{x}{\sqrt{1+x^{2}}}\) (see Table 3)  are carried out by the corresponding second-order ergodic mappings \(F_{2}(X)\) and 
\(F^{*}_{2}(X)\) at \(\alpha=1\).
In Fig. 8, we show that the superefficiency of chaos-based Monte Carlo computation 
is carried out for a two-dimensional integral of an integrand
 \begin{equation}
\label{eq:two_nong_integrand}
B(x,y)=P_{2}(x) P_{3}(y)-P_{1}(x)P_{1}(y)=\frac{(1-x^{2})(1-3y^{2})\mbox{sgn}(y)}{(x^{2}+1)(1+y^{2})^{\frac{3}{2}}}-
\frac{\mbox{sgn}(x)\mbox{sgn}(y)}{\sqrt{1+x^{2}}\sqrt{1+y^{2}}}
\end{equation}
(see also Table 3)
over the infinite square \(\Omega_{2}=(-\infty,+\infty)\times(-\infty,\infty)\) when we use
two different chaotic dynamical systems \(X_{n+1}=F_{2}(X_{n})\) and \(Y_{n+1}=F_{3}(Y_{n})\) at \(\alpha=1\) for the corresponding variables \(x\) and \(y\) of the two-dimensional integrand \(B(x,y)\).
Hence, it is shown that  chaos-based Monte Carlo computations 
can be performed to the integration problems over the infinite support by 
utilizing ergodic transformations over \(\Omega_{I}\) and superefficient Monte Carlo computation can also be attained if the superefficiency condition 
is satisfied as before. 
\section{Physical Meaning of Superefficiency}
Since the ergodic dynamical systems in this type of Monte Carlo computations 
generate Markov processes \cite{lasota}, 
we can  naturally construct stochastic processes 
obeying  Brownian motion from the dynamical systems.  
We consider a random variable  \(\delta B_{i}\equiv B_{i}-\langle B\rangle=B(X_{i})-\langle B(X_{i})\rangle \) generated  by a chaotic dynamical system
\(X_{n+1}=F(X_{n})\) as a velocity of a Brownian particle at \(t=i\).
 Here, the particle position of this discrete-time Brownian motion at \(t=N\)
is given by 
\begin{equation}
  r(N)\equiv \sum_{i=0}^{N-1}\delta B_{i}=\sum_{i=0}^{N-1}\{B_{i}-\langle B \rangle\}.
\end{equation}
The diffusion coefficient \(D_{discrete}\) is thus obtained by the formula 
\begin{equation}
\label{eq:discrete_diff}
D_{\mbox{discrete}}\equiv\lim_{N\rightarrow\infty}\frac{\langle\langle r^{2}(N)\rangle\rangle}{2N}=\lim_{N\rightarrow\infty}\{N\sigma(N)\}=
\frac{\langle B^{2}\rangle-\langle B\rangle^{2}}{2}
   +\sum_{j=1}^{\infty}\{\langle B_{0}B_{j}\rangle -\langle B\rangle^{2}\}.
\end{equation}
In the second equality we  used Eq. (\ref{eq:correlations}). Here, 
formula (\ref{eq:discrete_diff})  can thus be seen as   a discrete-time  version of the formula 
\begin{equation}
  D_{\mbox{continuous}}\equiv \lim_{t\rightarrow\infty} \frac{\langle\langle r(t)^{2}\rangle\rangle}{2t}=\int_{0}^{\infty}\langle u(0)u(t)\rangle dt,
\end{equation}
 for  continuous-time Brownian motion \(\{ r(t)\}\) with the particle velocity \(u(t)\) satisfying 
the  Langevin equation 
\begin{equation}
   \dot u(t)=-\gamma u(t)+R(t)
\end{equation}
where \(R(t)\) is assumed to be a Gaussian process such that 
\begin{equation}
  \langle\langle R(t)\rangle\rangle=0,\langle\langle R(t)R(t')\rangle\rangle =2\epsilon \delta(t-t')
\end{equation}
 and  \(r(t)=\int_{0}^{t}u(t')dt'\).  
Hence, superefficiency condition (\ref{eq:super_condition}) is simply 
rewritten as  
\begin{equation}
   D_{\mbox{discrete}}=\frac{\langle B^{2}\rangle-\langle B\rangle^{2}}{2}
   +\sum_{j=1}^{\infty}\{\langle B_{0}B_{j}\rangle -\langle B\rangle^{2}\}
=0.
\end{equation}
Note that \(D_{\mbox{discrete}}=0\) does not necessarily mean the corresponding discrete-time Brownian motion is dead, such as  Brownian motion at zero temperature. As is clearly seen in Fig. 9, the discrete-time Brownian motion which 
corresponds to the superefficiency \(D_{\mbox{discrete}}=0\) is still actively 
fluctuating around zero without diffusing while the discrete-time Brownian motion 
corresponding to the normal Monte Carlo simulations, \(D_{\mbox{discrete}}>0\), has a normal 
diffusion behavior such as \(\langle r^{2}(N)\rangle =O(N)\). We call such  non-diffusing  
Brownian motion {\it  ultracoherent Brownian motion}.
This ultracoherent Brownian motion (\(D_{\mbox{discrete}}=0\)) is quite unlike 
the zero diffusion  (\(D_{\mbox{continuous}}=0\)) case of continuous-time 
Brownian motion.  
Hence, the physical occurrence of the ultracoherent Brownian motion corresponding 
to the superefficient chaos Monte Carlo simulation is owing to the discrete 
nature of time in chaotic dynamical systems. 
Whether such ultracoherent Brownian motion is observed in  real physical systems would be 
an interesting problem beyond the scope of the present paper.

\section{Summary of the main results and conclusions} 
In this paper we show that Monte Carlo computation of a general 
 multi-dimensional integrals are performed by various ergodic 
dynamical systems utilized as non-uniform random-number generators. 
Such chaos-based Monte Carlo computations  
exhibit a dynamical effect of random-number generations that cannot be explained by the statistical arguments such as the importance sampling. By evaluating the expectation value \(\sigma(N)\) of the square of the error 
in terms of two-point correlation functions, the dynamical dependency of 
chaotic random-number generations is shown to cause a non-negligible effect 
in Monte Carlo simulations. While an effect of chaotic random-number generations 
is generally seen as a variance reduction effect caused by  negative covariance, 
we have shown  that a superefficient Monte Carlo computation can be carried out such  that the expectation value of the square of the error decreases to 0 as  \(\frac{1}{N^{2}}\) with \(N\) successive observations for 
\(N\rightarrow \infty\). Therefore,  
a proper choice of chaotic random-number 
generators  to cause superefficiency greatly accelerates
the speed of Monte Carlo computations. We have given the necessary and sufficient 
condition for achieving superefficient Monte Carlo computation. 
Interestingly, this superefficiency condition has  
 a physical meaning such that the diffusion coefficient of the corresponding discrete-time Brownian motion constructed from  chaotic random-number 
generators  vanishes (\(D_{discrete}=0\)), where the Brownian particles move actively in a localized region around the expectation value of the 
Monte Carlo simulation. We  numerically verified 
 the superefficiency of chaos Monte Carlo computations which 
sharply depends on the integrand and our choice of chaotic dynamical systems as 
random-number generators under various settings of  numerical integration problems.
Thus we can see that observed dynamical dependency resulting in variance 
reduction in Section 3 is a partial 
realization of an extreme case of dynamical dependency, i.e., {\it superefficiency}.

Monte Carlo methods are widely used in various 
computational problems (such as calculating financial derivative securities, 
communication traffic analysis, optimization problems, and biological computations.) Therefore, such remarkably  superior results from 
using  chaotic correlation 
here give us to hope that  they can  effectively be   
applied  to such hard computational problems. Furthermore, it is of special interest for us to  investigate an possibility that some 
physical (or biological) computations with unknown mechanism such as fast protein-folding computation, would utilize the present kinds of superefficient Monte Carlo computation  by 
the strong chaotic correlation of random-number generations. 

In conclusion, the present paper shows that 
 proper chaotic (ergodic)  random-number generators have a distinguishable 
effect from the conventional pseudo random-number generators such that  
chaotic random-number generators can
greatly accelerate the speed  of convergence of the mean square error 
in Monte Carlo computation (\(O(\frac{1}{N})\rightarrow O(\frac{1}{N^{2}})\)) 
by utilizing negative correlation.  Such superefficient Monte Carlo computation 
is robust under a change of the initial conditions, which  
is very promising for the application  of chaotic random-number generators to
 many other computational problems, in which the expectation value of the square 
of the error is usually considered to slowly converge to 0 as \(O(\frac{1}{N})\).\\[2cm]
{\bf Acknowledgements.}\\  
This work was initiated at the RIKEN Brain Science Institute. 
The author is grateful to S. Amari, Director of the Brain-Style Information Systems Research Group of the RIKEN Brain Science Institute for his generous support. 
The author would like to  thank S.  Amari, T. Khoda, T. Matsumoto, S. Tezuka, A. Yamaguchi, K. Goto, and T. Deguchi  for their valuable discussions. 
The author
 wishes to express his gratitude to  thank K. Kitayama, T. Itabe, and Y. Furuhama
 for their continuous encouragement.   
This work is, in part, supported by the  CRL Scientific Research Fund from the  
Ministry of Posts and Telecommunications. \\[1cm]
\appendix
\section{Derivation of Eq. (\ref{eq:correlations})}
In this appendix, we derive  formula (\ref{eq:correlations}).
\renewcommand\displayindent{5zw}
\begin{eqnarray*}
\sigma (N) & =& \langle\langle [\frac{1}{N}\sum_{j=0}^{N-1}B_{j}-\overline{B}]^{2}\rangle\rangle\\
 &  = & \langle\langle [\frac{1}{N}\sum_{j=0}^{N-1}B_{j}]^{2} \rangle\rangle
-2\langle\langle [\frac{1}{N}\sum_{j=0}^{N-1}B_{j}]\overline{B}\rangle\rangle +
\langle \langle\overline{B} \rangle\rangle^{2} \\
&  = & \frac{1}{N^{2}}\sum_{j=0}^{N-1}\langle B_{j}^{2}\rangle 
+\frac{2}{N^{2}}\sum_{j=1}^{N-1}(N-j)\langle B_{0}B_{j}\rangle -2
\langle \overline{B}^{2} \rangle +\langle B\rangle^{2}\\
& = & \frac{1}{N}\langle B^{2} \rangle +
\sum_{j=1}^{N}\frac{2(N-j)}{N^{2}}\langle B_{j}B_{0}\rangle -\langle B \rangle^{2}\\
& =& \frac{1}{N}\{\langle B^{2}\rangle -\langle B\rangle^{2}\}
    +\frac{2}{N^{2}}\sum_{j=1}^{N}(N-j)\{\langle B_{0}B_{j}\rangle -\langle B\rangle^{2}\}.
\end{eqnarray*}
In the third equality, we  used the invariance property:
\begin{equation}
   \langle B[F(X)]\rangle = \langle B \rangle
\end{equation}
and 
the following identities
\begin{equation}
   \langle\langle [\frac{1}{N}\sum_{j=0}^{N-1}B_{j}] \rangle\rangle =\frac{1}{N}\sum_{j=0}^{N-1}\langle\langle B_{j} \rangle\rangle = \langle B \rangle = \overline{B}.
\end{equation}
In the last equality, we  use the identity
\begin{equation}
  \frac{2}{N^{2}}\sum_{j=1}^{N}(N-j)\langle B\rangle^{2}=\langle B\rangle^{2}-\frac{1}{N}\langle B\rangle^{2}. 
\end{equation}

\section{Derivation of Eq. (\ref{eq:s_dimensional_sigma})}
 In this appendix, we derive  formula (\ref{eq:s_dimensional_sigma}).

Since  the following equalities
\begin{eqnarray*}
2\sum_{l=1}^{N}(1-\frac{l}{N})\{\langle B_{0}B_{l} \rangle-\langle B\rangle^{2}\} & = & \frac{1}{2^{s-1}}\sum_{l=1}^{L}(1-\frac{l}{N})\sum_{m=l}^{L}a_{m}a_{m-l}\\
&=& \frac{1}{2^{s}}\{(\sum_{m=0}^{L}a_{m})^{2}-(\sum_{m=0}^{L}a_{m}^{2})\}
-\frac{1}{N2^{s-1}}\sum_{l=1}^{L}\sum_{m=l}^{L}a_{m}a_{m-l}l,
\end{eqnarray*}
and 
\begin{eqnarray*}
  \langle B^{2}\rangle -\langle B \rangle^{2}=\frac{1}{2^{s}}\sum_{m=0}^{L}a_{m}^{2},
\end{eqnarray*}
hold,
 we have for \(N\geq L\)
\begin{eqnarray*}
\sigma(N)& =& \frac{1}{N}\{\langle B^{2}\rangle -\langle B\rangle^{2}\}
 +\frac{2}{N}\sum_{l=1}^{N}(1-\frac{l}{N})\{\langle B_{0}B_{l}\rangle-\langle B\rangle^{2}\} \\
 & = & \frac{1}{N}\frac{\sum_{m=0}^{L}a_{m}^{2}}{2^{s}}+\frac{1}{N}
       \frac{\{(\sum_{m=0}^{L}a_{m})^{2}-\sum_{m=0}^{L}a_{m}^{2}\}}{2^{s}}-
       \frac{1}{2^{s-1}N^{2}}\sum_{l=1}^{L}\sum_{m=l}^{L}a_{m}a_{m-l}l \\
 & = & \frac{1}{N}\frac{(\sum_{m=0}^{L}a_{m})^{2}}{2^{s}}-\frac{1}{2^{s-1}N^{2}}
    \sum_{l=1}^{L}\sum_{m=l}^{L}a_{m}a_{m-l}l.
\end{eqnarray*}

\section{Orthonormal System of Functions Related to the Topological Conjugacy Relation with  Chebyshev Maps}
In this appendix, we derive various 
orthonormal functions  from  ergodic mappings by using 
the topological conjugacy relations with 
Chebyshev polynomials. Let us consider the topological conjugacy 
relation with the Chebyshev maps \(f\in \{T_{m}\}_{m=0,\cdots}\)
\begin{equation}
   \tilde f(y) = h^{-1}\circ f \circ h(y),  
\end{equation}
 where \(\tilde f\) and \(f\) are onto-mappings such that 
\(\tilde f:\Omega_{I}\rightarrow \Omega_{I}\) and  \(f: \Omega\rightarrow \Omega\), and 
 \(h\) is a differentiable one-to-one mapping (diffeomorphism) as \(h:\Omega_{I}\rightarrow \Omega\). Since \(h\) is a diffeomorphism, \(h\) preserves the ergodicity, and the mixing property of \(f\); \(\tilde f\) also has  a mixing (thus, ergodic) property 
if \(f\) has the mixing (thus, ergodic) property.
Recall that the \(p\)-th Chebyshev maps at \(p\geq 2\) have
the invariant measures \(\rho(x)dx=\frac{dx}{\sqrt{1-x^{2}}}\) and they satisfy
the orthogonal relation with respect to this measure \(\rho(x)dx\). Suppose that 
the mapping \(\tilde f\) has the invariant probability measure \(\rho_{NG}(y)dy\).
Then, the 
 probability preservation relation 
\begin{equation}
   \rho(x)|dx|=\rho_{NG}(y)|dy|
\end{equation} 
holds. 
We thus obtain the density function \(\rho_{NG}(y)\) for \(\tilde f\) as follows:
\begin{equation}
\label{eq:g_nong_density}
  \rho_{NG}(y)=\rho[h(y)]|\frac{dh(y)}{dy}|.
\end{equation} 
Here  we consider a system of mappings \(\{F_{l}\}_{l=0,\cdots,\infty}\) where \(F_{l}\) is defined by the topological conjugacy with the \(l\)-th order 
Chebyshev map
\begin{equation}
  F_{l}(y)=h^{-1}\circ T_{l}\circ h (y).
\end{equation}
Since the Chebyshev maps \(\{ T_{l}\}\) at \(l\geq 2\) have the mixing property, all of the corresponding maps
\(\{F_{l}\}\) at \(\geq 2\) also have  the mixing property. 
We therefore have the following theorem.
\begin{theorem}
\label{theorem :A1}
A systems of functions \(\{ P_{l}(y)\}_{l=0,\cdots,\infty}\) defined as 
\begin{eqnarray}
\nonumber
    P_{l}(y)\equiv T_{l}[h(y)]=h^{-1}[F_{l}(y)]
\end{eqnarray}
constitutes an orthonormal system of functions such that  
\begin{equation}
\label{eq:nong_ort}
\int_{\Omega_{I}} P_{i}(y) P_{j}(y) \rho_{NG}(y)dy=
 \delta_{i,j}\frac{(1+\delta_{i,0})}{2},
\end{equation}
where \(\rho_{NG}(y)\) is a density function defined in 
Eq. (\ref{eq:g_nong_density}).
\end{theorem}
The proof of the theorem is given  by the following identity
\renewcommand\displayindent{3zw}
\begin{eqnarray}
 \int_{\Omega_{I}} P_{i}(y) P_{j}(y)\rho_{NG}(y)dy & 
= & \int_{ \Omega}h\circ F_{i}\circ h^{-1}(x)\cdot h\circ F_{j}\circ 
h^{-1}(x)\rho(x)dx \nonumber \\ 
  & 
= & \int_{ \Omega} T_{i}(x)\cdot  T_{j}(x)\rho(x)dx \\
 & 
= & \delta_{i,j}\frac{(1+\delta_{i,0})}{2} \nonumber.
\end{eqnarray}
In the last equality, we use the orthonormal property of the Chebyshev system
in Eq. (\ref{eq:che_orthonormal}). The topological conjugacy relation with the Chebyshev maps thus gives
the corresponding orthonormal system \(\{ P_{l}(y)\}\) with the density 
function \(\rho_{NG}(y)\) as well as 
the corresponding ergodic mapping \(F_{l}(y)\) with the invariant probability 
measure \(\rho_{NG}(y)dy\). 
\section{Non-Gaussian Chaos Mappings on the Real Line}
In this appendix, we derive various ergodic transformations on the real line \((-\infty,\infty)\) via the topological conjugacy relations with the Chebyshev 
mappings \(T_{l}(x)\). Let us consider a diffeomorphism 
\(h:(-\infty,\infty)\rightarrow (-1,1)\) given by  
\begin{equation}
  h(y)=\frac{1}{\sqrt{1+|y|^{2\alpha}}}\mbox{sgn}(y)\quad \mbox{for}\quad \alpha>0.
\end{equation} 
The topological conjugacy relation with the Chebyshev mappings \(T_{l}\)
 gives a  set of infinite numbers of ergodic transformations \(F_{l}(y)=h^{-1}\circ T_{l}\circ h(y)\) with the invariant measure \( \rho_{NG}(y)dy\), where the density \(\rho_{NG}(y)\)is given by the formula
\begin{equation}
\label{eq:nong_density}
   \rho_{NG}(y)=\rho[h(y)]|\frac{h(y)}{dy}|=
\frac{\alpha|y|^{\alpha-1}}{\pi(1+|y|^{2\alpha})}. 
\end{equation}
Note that when \(\alpha=1\), the density function (\ref{eq:nong_density}) corresponds to the Cauchy density function \( \rho_{NG}(y)=\frac{1}{\pi(1+y^{2})}\)
 with  infinite 
variance. 
In general,  density function (\ref{eq:nong_density}) for \(<\alpha <2\) 
does not have a finite variance. 
As a result, the central limit theorem is not applicable here and then 
a superposition of random variables obeying this density 
does not converge in distribution to the Gaussian distribution  but converge to the {\it non-Gaussian L\'evy's stable} distribution\cite{gnedenko}. Thus, we call these dynamical systems 
{\it non-Gaussian chaos}. Here, we consider 
a function \(f(x)\) as an integrand in  the  functional  space \(\cal L\mit^{2}(\Omega_{I},\rho_{NG})\) such as    
\begin{equation}
\cal L\mit^{2}(\Omega_{I},\rho_{NG})
=\{f(x)|  \int_{\Omega_{I}}|f(x)|^{2}\rho_{NG}(x)dx<\infty\},
\end{equation} 
which are spanned by the  orthonormal system of functions \(\{ P_{l}(y)\}\).
The functional space \(\cal L\mit^{2}(\Omega_{I},\rho_{NG})\) for \(0<\alpha<2\) 
is clearly different from the functional space 
\(\cal L\mit^{2}(\Omega_{I},e^{-x^{2}})\)
which 
can be spanned by the complete orthonormal system of the Hermite polynomials \(H_{l}(x)=(-1)^{l}e^{x^{2}}\frac{d^{l}}{dx^{l}}e^{-x^{2}}\) with respect to the Gaussian measure \(e^{-x^{2}}dx\).   
According to Ref. \cite{ku9}, we can consider an ergodic 
transformation \(F_{l}(y)\) with the density (\ref{eq:nong_density})
as the multiplication formula of \(|\tan(\theta)|^{\frac{1}{\alpha}}\mbox{sgn}[\cos(\theta)]\), that is, 
\begin{equation}
    F_{l}(|\tan(\theta)|^{\frac{1}{\alpha}}\mbox{sgn}[\cos(\theta)])=
    |\tan(l\theta)|^{\frac{1}{\alpha}}  \mbox{sgn}[\cos(l\theta)].
\end{equation}
In this way, we can easily obtain a series of ergodic transformations \(\{F_{l}\}_{l=2,3,\cdots}\)
with non-Gaussian density function over the infinite support \((-\infty,\infty)\) as follows:
\begin{equation}
 F_{2}(y)=|\frac{2|y|^{\alpha}}{1-|y|^{2\alpha}}|^{\frac{1}{\alpha}}
\mbox{sgn}(-|y|^{2\alpha}+1),F_{3}(y)=|\frac{|y|^{\alpha}(|y|^{2\alpha}-3)}{1-3|y|^{2\alpha}}|^{\frac{1}{\alpha}}\mbox{sgn}[y(1-3|y|^{2\alpha})],\cdots.
\end{equation}
In the same way as used in the preceding appendix, orthogonal functions 
\(\{ P_{l}(y)\}\) satisfying Eq. (\ref{eq:nong_ort})
with respect to the density \(\rho_{NG}(y)\) 
over the infinite support \(\Omega_{I}=(-\infty,\infty)\)
are derived as follows:
\begin{equation}
 P_{0}(y)=1, P_{1}(y)=h(y),
 P_{2}(y)=\frac{-|y|^{2\alpha}+1}{|y|^{2\alpha}+1},
 P_{3}(y)=\frac{1-3|y|^{2\alpha}}{(1+|y|^{2\alpha})^{\frac{3}{2}}}\mbox{sgn}(y),\cdots.
\end{equation}
On the contrary, 
if we choose a diffeomorphism \(h^{*}:\Omega_{I}\rightarrow \Omega\) given by 
\begin{equation}
    h^{*}(y)=\frac{|y|^{\alpha}}{\sqrt{1+|y|^{2\alpha}}}\cdot \mbox{sgn}(y),
\end{equation}
the corresponding ergodic transformations \(F^{*}_{l}(y)\) are 
derived from the Chebyshev mappings \(T_{l}(y)\) via the topological conjugacy relation 
\begin{equation}
h^{*}\circ F_{l}^{*}(y)= T_{l}\circ h^{*}(y).
\end{equation}
In this case, \(F^{*}_{l}\) is equivalent to a multiplication formula of 
\(|\cot (\theta)|^{\frac{1}{\alpha}}\mbox{sgn}[\cos(\theta)]\) as 
\begin{equation}
F^{*}_{l}(|\cot (\theta)|^{\frac{1}{\alpha}}\mbox{sgn}[\cos(\theta)])
 =|\cot (l\theta)|^{\frac{1}{\alpha}}\mbox{sgn}[\cos(l\theta)].
\end{equation}
We thus obtain a series of ergodic transformations \(\{F^{*}_{l}(y)\}\) 
such that 
\begin{equation}
  F^{*}_{2}(y)=|\frac{1}{2}(|y|^{\alpha}-\frac{1}{|y|^{\alpha}})|^{\frac{1}{\alpha}}
 \mbox{sgn}(|y|-\frac{1}{|y|}), F^{*}_{3}(y)=
|\frac{|y|^{\alpha}(3-|y|^{2\alpha})}{1-3|y|^{2\alpha}}|^{\frac{1}{\alpha}}\mbox{sgn}[y(|y|^{2\alpha}-3)],\cdots.
\end{equation}
Correspondingly, orthogonal functions  \(\{ P^{*}_{l}(y)\}\)  related to the  transformations \(\{F^{*}_{l}(y)\}_{l=0,\cdots}\) are derived as follows:
\begin{equation}
  P^{*}_{0}(y)=1, P^{*}_{1}(y)=h^{*}(y),
P^{*}_{2}(y)=\frac{|y|^{2\alpha}-1}{|y|^{2\alpha}+1},
P^{*}_{3}(y)=\frac{|y|^{\alpha}(|y|^{2\alpha}-3)}{(1+|y|^{2\alpha})^{\frac{3}{2}}}
\mbox{sgn}(y),\cdots.
\end{equation}
In the construction of the above ergodic  mappings, there exists a duality   such that 
\begin{equation}
    F_{l}(y) F^{*}_{l}(z)=1,\quad \mbox{for}\quad yz=1.
\end{equation}
Table 3 summarizes the analytical results here 
for the ergodic transformations and the related orthogonal  functions, which 
are necessary  
for producing 
the numerical results of the superefficient chaos Monte Carlo computations
over the infinite support \(\Omega_{I}\) in Section 7. The graphs of ergodic transformations 
over \(\Omega_{I}\) are depicted in Fig. 6(a).

\clearpage
\begin{center}
{\bf Table 1: Ergodic Mappings with Lyapunov exponents \(\lambda\) on [0,1]}
\end{center}
\begin{tabular}{|l|c|c|}
\hline \hline {\bf Rational Mappings on [0,1]} & 
{\bf Densities \(\rho(x)\)} & \(\lambda\) \\ \hline
{\bf \(f^{(2)}(x)=4x(1-x)\)} (Ulam-von Neumann Map)\cite{ulam} & 
   \(\frac{1}{\pi\sqrt{x(1-x)}}\) & \(\mbox{ln}2\) \\ \hline
{\bf  \(f^{(3)}(x)=x(3-4x)^{2}\)} (Cubic Map)\cite{adler} & 
\(\frac{1}{\pi\sqrt{x(1-x)}}\) & \(\mbox{ln}3\) \\ \hline
{\bf \(f^{(5)}(x)=x(5-20x+16x^{2})^{2}\)}(Quintic Map)\cite{adler} &
\(\frac{1}{\pi\sqrt{x(1-x)}}\) & \(\mbox{ln}5\) \\  \hline
{\bf \(f^{(2)}_{l,m}(x)=\frac{4x(1-x)(1-lx)(1-mx)}{1-2(l+m+lm)x^{2}+
8lmx^{3}+(l^{2}+m^{2}-2lm-2l^{2}m-2lm^{2}+l^{2}m^{2})x^{4}}\)}\cite{ku7} & 
\(\frac{1}{K\sqrt{x(1-x)(1-lx)(1-mx)}}\) & \(\mbox{ln}2\) \\ \hline
{\bf \(f^{(3)}_{l,m}(x)= [x\lbrace-3+4x+4(l+m)x-6(l+m+lm)x^{2}\) } & \mbox{ } & \mbox{ } \\ 
{\bf \(+12lmx^{3}+(l^{2}+m^{2}-2lm-2l^{2}m-2lm^{2}+l^{2}m^{2})x^{4}\rbrace^{2}]/ \)} & 
\(\frac{1}{K\sqrt{x(1-x)(1-lx)(1-mx)}}\) & \(\mbox{ln}3\) \\ 
{\bf \([ 1-12(l+m+lm)x^{2}+8(l+m+l^{2}+m^{2}+l^{2}m+lm^{2})x^{3}\)} &
\mbox{ } &  
\mbox{  }\\ 
{\bf \(+120lmx^{3}+6(5l^{2}+5m^{2}-26lm-26l^{2}m-26lm^{2}+5l^{2}m^{2})x^{4}\)} & \mbox{ } & \mbox{  }\\ 
{\bf \(+24(-2l^{2}-2m^{2}-2l^{3}-2m^{3}+4lm+7l^{2}m+7lm^{2})x^{5}\)} & \mbox{ } & \mbox{  }\\ 
{\bf \(+24(4l^{3}m+4lm^{3}+7l^{2}m^{2}-2l^{3}m^{2}-2l^{2}m^{3})x^{5}\)} & 
\mbox{  } & \mbox{  }\\ 
{\bf \(+4(4l^{2}+4m^{2}+4l^{4}+4m^{4}+17l^{3}+17m^{3}-8lm)x^{6}\)} & \mbox{ } & \mbox{  }\\ 
{\bf \(+4(-17l^{2}m-17lm^{2}-17l^{3}m-17lm^{3}-8l^{4}m-8lm^{4})x^{6}\)} & \mbox{ } & \mbox{  }\\ 
{\bf \(+4(4l^{2}m^{4}+4l^{4}m^{2}-17l^{3}m^{2}-17l^{2}m^{3}+17l^{3}m^{3}-
54l^{2}m^{2})x^{6}\)} & \mbox{ } & \mbox{  }\\ 
{\bf \(+24(-l^{3}-m^{3}-l^{4}-m^{4}+l^{2}m+lm^{2}-l^{3}m-lm^{3})x^{7}\)} & 
\mbox{ } & \mbox{  }\\  
{\bf \(+24(l^{4}m+lm^{4}+4l^{2}m^{2}+4l^{3}m^{2}+4l^{2}m^{3})x^{7}\)} & 
\mbox{ } & \mbox{  }\\  
{\bf \(+24(l^{4}m^{2}+l^{2}m^{4}-l^{3}m^{3}-l^{4}m^{3}-l^{3}m^{4})x^{7}\)} & 
\mbox{ } & \mbox{  }\\ 
{\bf \(+3(3l^{4}+3m^{4}+4l^{3}m+4lm^{3}+4l^{4}m+4lm^{4}-14l^{2}m^{2})x^{8}\)} & 
\mbox{ } & \mbox{  }\\ 
{\bf \(+3(-4l^{3}m^{2}-4l^{2}m^{3}-4l^{3}m^{3}-14l^{4}m^{2}-14l^{2}m^{4})x^{8}\)} & \mbox{ } & \mbox{  }\\ 
{\bf \(+3(4l^{4}m^{3}+4l^{3}m^{4}+3l^{4}m^{4})x^{8}\)} & \mbox{ } & \mbox{  }\\ 
{\bf \(+8(-l^{4}m-lm^{4}+l^{3}m^{2}+l^{2}m^{3}+l^{4}m^{2}+l^{2}m^{4})x^{9}\)} & \mbox{ } & \mbox{  }\\ 
{\bf \(+8(-2l^{3}m^{3}+l^{4}m^{3}+l^{3}m^{4}-l^{4}m^{4})x^{9}] \)} \cite{ku7} & \mbox{ } & \mbox{  } \\ \hline
\end{tabular}
\clearpage
\begin{center}
{\bf Table 2: Convergence speed of various types of Monte Carlo simulations  with \(s\)-dimensional integration problems}\\[0.2cm]
\begin{tabular}{|l|c|}
\hline
\hline {\bf Random-Number Generators}
     &  {\bf Variance of  Error} \\ \hline
Standard Arithmetical Pseudo-Random Numbers (General) &  \(V(N)= O(\frac{1}{N})\) \\ \hline
Quasi-Random Numbers (General)& \(V(N)=O[\frac{1}{N^{2}}\cdot(\mbox{ln}N)^{2s}]\) \\ \hline 
Superefficient Chaos Monte Carlo [under  Condition (\ref{eq:super_condition})] &  \(V(N)=O(\frac{1}{N^{2}})\) \\ \hline
Chaos Monte Carlo (General) & \(V(N)=O(\frac{1}{N})\) \\
\hline
\end{tabular} 
\end{center}
\clearpage
\begin{center}
{\bf Table 3: Orthogonal functions related to exactly solvable chaos   }\\[0.2cm]
\begin{tabular}{|l|c|c|c|}
\hline
\hline {\bf Index}
     &  {\bf \(P_{l}(x)\) } & {\bf \(F_{l}(x)\)} & Lyapunov Exponents 
\\ \hline
\(l=1\) &  \(\frac{1}{\sqrt{1+|x|^{2\alpha}}}\mbox{sgn}(x)\) &
\(x\) & \(0\) \\ \hline 
\(l=1\)*(Dual) & 
\(\frac{|x|^{\alpha}}{\sqrt{1+|x|^{2\alpha}}}\mbox{sgn}(x)\) & \(x\) & \(0\) \\ \hline
\(l=2\) & \(\frac{-|x|^{2\alpha}+1}{|x|^{2\alpha}+1}\) & 
\(|\frac{2|x|^{\alpha}}{1-|x|^{2\alpha}}|^{\frac{1}{\alpha}}
\mbox{sgn}(-|x|^{2\alpha}+1)\) & \(\mbox{ln}2\) \\ \hline 
\(l=2\)*(Dual) &  \(\frac{|x|^{2\alpha}-1}{|x|^{2\alpha}+1}\) 
 & \(|\frac{1}{2}(|x|^{\alpha}-\frac{1}{|x|^{\alpha}})|^{\frac{1}{\alpha}}
 \mbox{sgn}(|x|-\frac{1}{|x|})\) & \(\mbox{ln}2\)
\\ \hline 
\(l=3\) &  \(\frac{1-3|x|^{2\alpha}}{(1+|x|^{2\alpha})^{\frac{3}{2}}}\mbox{sgn}(x)\) & 
\(|\frac{|x|^{\alpha}(|x|^{2\alpha}-3)}{1-3|x|^{2\alpha}}|^{\frac{1}{\alpha}}\mbox{sgn}[x(1-3|x|^{2\alpha})]\) 
& \(\mbox{ln}3\) \\ \hline
\(l=3\)*(Dual) &  
\(\frac{|x|^{\alpha}(|x|^{2\alpha}-3)}{(1+|x|^{2\alpha})^{\frac{3}{2}}}
\mbox{sgn}(x)\) & \(|\frac{|x|^{\alpha}(3-|x|^{2\alpha})}{1-3|x|^{2\alpha}}|^{\frac{1}{\alpha}}\mbox{sgn}[x(|x|^{2\alpha}-3)]\) & \(\mbox{ln}3\)
\\
\hline
\end{tabular} 
\end{center}
\end{document}